\documentclass[preprintnumbers,superscriptaddress,showkeys,showpacs,byrevtex]{revtex4}

\usepackage{amsmath,amsfonts,amssymb,amscd,amsxtra,amsthm}
\usepackage{graphicx}
\usepackage{epstopdf}
\usepackage{bm}
\usepackage{feynmp}
\usepackage{feynmp-auto}
\usepackage[colorlinks=true, pdfstartview=FitV, bookmarks=true, bookmarksnumbered=true, breaklinks]{hyperref}
\usepackage{mathtools,braket}
\usepackage{soul}
\usepackage{lipsum} 
\usepackage{slashed}
\usepackage{xcolor}
\usepackage{physics}
\usepackage{multirow}

\usepackage{newtxmath}
\usepackage{mathrsfs}

\definecolor{blue}{rgb}{0.0, 0.0, 1.0}
\definecolor{red}{rgb}{1.0, 0.0, 0.0}
\definecolor{royalblue}{rgb}{0.0, 0.14, 0.4}
\hypersetup{linkcolor=royalblue, citecolor=blue, urlcolor=royalblue}

\usepackage{hyperref}
\hypersetup{colorlinks=true,citecolor=blue,linkcolor=blue,urlcolor=blue}

\usepackage[mathlines]{lineno}
\usepackage{tikz,xcolor}
\definecolor{lime}{HTML}{A6CE39}
\DeclareRobustCommand{\orcidicon}{%
	\begin{tikzpicture}
	\draw[lime, fill=lime] (0,0) 
	circle [radius=0.16] 
	node[white] {{\fontfamily{qag}\selectfont \tiny ID}};
	\draw[white, fill=white] (-0.0625,0.095) 
	circle [radius=0.007];
	\end{tikzpicture}
	\hspace{-2mm}
}

\foreach \x in {A, ..., Z}{%
	\expandafter\xdef\csname orcid\x\endcsname{\noexpand\href{https://orcid.org/\csname orcidauthor\x\endcsname}{\noexpand\orcidicon}}
}

\begin{document}
%\preprint{PKNU-NuHaTh-20XX-XX}
\title{Antineutrino Opacity in Neutron Stars in the Models Constrained by Recent Terrestrial Experiments and Astrophysical Observations}
%--------------------------------------------------
\author{Parada~T.~P.~Hutauruk\orcidB{}}
\email{phutauruk@pknu.ac.kr}
\affiliation{Department of Physics, Pukyong National University (PKNU), Busan 48513, Korea}
%--------------------------------------------------
\date{\today}
\begin{abstract}
In the present paper, I have investigated a neutral current (NC) antineutrinos scattering with neutron star (NS) matter constituents at zero temperature. The modeling of the standard matters in NS is constructed in the framework of both extended relativistic mean-field (E-RMF) and nonrelativistic Korea-IBS-Daegu-SKKU energy density functional (KIDS-EDF) models. In the E-RMF model, I use a new parameter of G3(M), which was constrained by the recent PREX II experiment measurement of neutron distribution of $^{208}\rm{Pb}$, while the  KIDS-EDF models are constrained by terrestrial experiments, gravitational-wave signals, and astrophysical observations. Using both realistic and well-constrained matter models, I then calculate the antineutrino differential cross-section (ADCS) and antineutrino mean free path (AMFP) of the antineutrinos-NS matter constituents interaction using a linear response theory. It is found that the AMFP for the KIDS0 and KIDSA models are smaller in comparison to the SLy4 model and the E-RMF model with the G3(M) parameter. The AMFP result of the Skyrme model with the SLy4 parameter set is found to have an almost similar prediction to that of the E-RMF model with the G3(M) parameter. Contributions of each nucleon to total AMFP are also presented for the G3(M) model.
\end{abstract}
\maketitle
%--------------------------------------------------
\section{Introduction}
%--------------------------------------------------
Neutrino emission is also one of the sources of the multi-messengers of astronomy which opens a great opportunity to provide fundamental information for the stellar object and compact star properties, besides a gravitational wave that was first detected by the Laser Interferometer Gravitational-Wave Observatory LIGO/Virgo Collaboration \textit{via} gravitational waveform generated in the merger of the binary neutron stars (NSs)~\cite{LIGOScientific:2017vwq}, cosmic rays, and electromagnetic radiation. Even though neutrinos weakly interact with matter, the giant detector has been successful in detecting a few astrophysical neutrinos. For example, a burst of the neutrino generated from Supernova SN1987A was detected in Superkamiokonde II, which was the first evidence of direct observation of neutrino astronomy~\cite{Hirata:1988ad,Burrows:1987zz}. Since then neutrino emissions have been detected from the astrophysical flux objects with high energy. A first observation corresponding to the high-energy astrophysical neutrino was recently reported by IceCube~\cite{IceCube:2016tpw}. However, the origin of neutrino emission is not yet fully understood. More improvements in detectors and analyses are needed to answer this question. 

In supernova explosions, many neutrinos are produced, where the white dwarfs from a pre-supernova star collapse into a protoneutron star (PNS). In this explosion process, the neutrinos are expected to diffuse out from the neutron-rich matter (NM) of the PNS. During the time of the diffusion from the PNS, a few percent of neutrinos interact with the matter outside the PNS. The interaction of neutrinos with matter can happen through the neutral-current scattering or the charged-current absorption. Instead of neutrinos, antineutrinos also can be produced during the supernova explosion. It was found that the DCS of the antineutrinos would be considerably smaller than that of the neutrinos~\cite{Horowitz:2003yx}. This implies the mean free path of the antineutrinos is considerably larger than that for neutrinos. This happens if the weak magnetism of the nucleon is considered in the calculation. This finding was also supported by Hutauruk, \textit{et al,} in Ref.~\cite{Hutauruk:2006re}, where the weak magnetism and neutrino form factors are included in the calculation. Motivated by these studies, in the present work, I focus on antineutrino's neutral-current scattering using the well-constrained matter model by terrestrial experiments and astrophysical observations data.

To realistically model the NS matter, it is important to care for the matter model's stability. The model must reproduce a binding energy per nucleon $E_B/A =$ 15 MeV at a saturation density $n_0 =$ 0.16 fm$^{-3}$. In E-RMF model, coupling constants are dictated by fitting the binding energy per nucleon at the saturation density as well as reproducing the neutron skin thickness that was measured in the PREX II experiment~\cite{PREX:2021umo}. Details of fitting strategy in extracting the coupling constants in the E-RMF model are referred to Ref.~\cite{Pattnaik:2021ido,Hutauruk:2021cgi}. The fitting strategy in the KIDS-EDF model is rather different from that in the E-RMF model. In the EDF model, the unknown parameters are matched with the conventional Skyrme force parameters, where the Skyrme force parameters are extracted by reproducing the skin thickness or charge radius that was measured in the terrestrial experiments and reproducing the NS radius or mass which was taken from the astrophysical observations such as the LIGO/VIRGO~\cite{LIGOScientific:2017vwq,LIGOScientific:2018cki}, and the Neutron Star Interior Composition Explorer (NICER)~\cite{Miller:2019cac,Riley:2019yda}. Details of a fitting strategy of the KIDS-EDF model can be found in Ref.~\cite{Papakonstantinou:2016zpe}. It is worth noting that a well-constrained EoS and antineutrino/neutrino mean free path ((A)NMFP) are really needed for a supernova simulation as inputs~\cite{Mezzacappa:2000jb,Rampp:2000ws}.

In this paper, I compute the DCS of antineutrinos and AMFP in terms of the neutral current scattering in the framework of the E-RMF and KIDS-EDF models~\cite{Pattnaik:2021ido,Hutauruk:2021cgi,Papakonstantinou:2016zpe}. Before calculating the DCS and AMFP in the scattering, I analyze the matter model that will be used as inputs to calculate the DCS and AMFP. To do that, I evaluate the EoS of the E-RMF and KIDS-EDF models, where the coupling constants of the models are constrained by the terrestrial laboratory experiments and the astrophysical observations, to guarantee the stability of the matter. Also, the EoS must reproduce the binding energy per nucleon at the saturation density. Thus, I performed a calculation of ADCS and AMFP for both E-RMF and KIDS-EDF models. A prediction result of the AMFP for the KIDS0 model shows that it decreases as the $n_B/n_0$ increases. This result is followed by the KIDSA and SLy4 models. However, starting at around $n_B/n_0 =$ 3.5, it is found that the AMFP of the KIDS0 model is lower than that of the KIDSA and SLy4 models. This can be understood due to the KIDS0 model having a larger effective nucleon mass at that baryon number density. Similarly, it is also shown that the AMFP for the E-RMF model with the G3(M) parameter decreases as baryon density increases. In addition, I also found that the AMFP of the E-RMF model with the G3(M) parameter slowly decreases in comparison with those of the KIDS0, KIDSA, and SLy4 models.

The present paper is organized as follows. In Sec.~\ref{sec:theor}, I briefly introduce the theoretical framework of the model of the nuclear matter that is constructed in the E-RMF and KIDS-EDF models. Results for the EoS, effective nucleon mass, and particle fraction for both models are also presented. In Sec.~\ref{sec:scatt}, I describe the neutrino-NS constituents scattering, which is constructed from the effective Lagrangian. The final expression for the antineutrino and neutrino DCS and MFP are presented. Section~\ref{sec:result} presents the discussions of DCS of antineutrino and the AMFP numerical results for both E-RMF and KIDS-EDF models. A summary and conclusion are given in Sec.~\ref{sec:summary}.

%--------------------------------------------------
\section{Theoretical Framework}
\label{sec:theor}
%--------------------------------------------------
In this section, I present the EoS of the nuclear matter model that describes the constituents matter of NS. The E-RMF model~\cite{Furnstahl:1995zb,Furnstahl:1996wv}, inspired by the effective field theory, with the modified G3(M) parameter is employed to describe the NS matter. This model has successfully been applied in various phenomena of nuclear physics~\cite{Hutauruk:2023mjj,Hutauruk:2004uf}. Besides the E-RMF model, the nonrelativistic EDF models have also been used to describe the NS matter. One of them is the KIDS-EDF model, which was developed first time in the homogenous NM~\cite{Papakonstantinou:2016zpe}. The KIDS-EDF model was constructed in terms of the expansion of the Fermi-momentum. In fact, the E-RMF and KIDS-EDF models are principally constructed on a different basis. Therefore, these two different models are opted to be used in the present work.

%-------------------------------------------------------------------
\subsection{Modeling of Nuclear Matter}
%-------------------------------------------------------------------
Here, the description of the NS matter is modeled in the E-RMF and KIDS-EDF models. Again, we emphasize that both models reproduce binding energy $E_B/A =$ 15 MeV at the saturation density $n_0 =$ 0.16 fm$^{-3}$ to fulfill the matter stability condition. A detailed description of both models is explained in Sec.~\ref{sec:RMF} and Sec.~\ref{sec:KIDSEDF}.

%-------------------------------------------------------------------
\subsubsection{E-RMF model}
\label{sec:RMF}
%-------------------------------------------------------------------
In the E-RMF model, commonly, the interactions among the nucleons and mesons, the mesons self-interacting, and mesons-mesons crossing interactions can be constructed in the effective Lagrangian. The expression of the effective Lagrangian for the E-RMF model is compactly written as~\cite{Furnstahl:1995zb,Furnstahl:1996wv}
\begin{eqnarray}
  \label{eq:eqxenont1}
  \mathcal{L}_{\rm{ERMF}} &=& \mathcal{L}_{\rm NM} + \mathcal{L}_{ \sigma}
  + \mathcal{L}_{ \omega} + \mathcal{L}_{ \rho}
  + \mathcal{L}_{ \delta}
  + \mathcal{L}_{ \sigma \omega \rho},  
\end{eqnarray}
where the first term of the effective interaction Lagrangian of the nucleons and mesons is given by
\begin{eqnarray}
  \label{eq:eqxenont2}
  \mathcal{L}_{\rm NM} &=& \sum_{j=n,p} \bar{\psi}_j \left[ i \gamma^\mu \partial_\mu
    - (M - g_\sigma \sigma - g_\delta \bm{\tau}_j \cdot \bm{\delta}) \right. 
    - \left( g_\omega \gamma^\mu \omega_\mu
    + \frac{1}{2} g_\rho \gamma^\mu \bm{\tau}_j \cdot \bm{\rho}_\mu \right) \psi_j ,
\end{eqnarray}
where $M$ is the nucleon mass in free space (vacuum), and $\bm{\tau}_j$ is the isospin matrices. The total interaction effective Lagrangian is for protons and neutrons, which are given by a sum mathematical symbol. The coupling constants of $g_\sigma$, $g_\omega$, $g_\rho$, and $g_\delta$ stand for the $\sigma$, $\omega$, $\rho$, and $\delta$ mesons, respectively. The second, third, fourth, and fifth terms of the self-interactions Lagrangian for the $\sigma$, $\omega$, $\rho$, and $\delta$ mesons are respectively defined by
\begin{eqnarray}
  \label{eq:eqxenon3}
  \mathcal{L}_{\sigma} &=& \frac{1}{2} \left( \partial_\mu \sigma \partial^\mu \sigma -m_\sigma^2 \sigma^2 \right)
  - \frac{\kappa_3}{6M} g_\sigma m_\sigma^2 \sigma^3 
  - \frac{\kappa_4}{24 M^2} g_\sigma^2 m_\sigma^2 \sigma^4 , \\
  \mathcal{L}_{\omega} &=& - \frac{1}{4} \omega_{\mu \nu} \omega^{\mu \nu} + \frac{1}{2} m_\omega^2 \omega_\mu \omega^\mu
  + \frac{1}{24} \xi_0 g_\omega^2 (\omega_\mu \omega^\mu)^2 , \\
  \mathcal{L}_\rho &=& -\frac{1}{4} \bm{\rho}_{\mu \nu} \cdot \bm{\rho}^{\mu \nu} + \frac{1}{2} m_\rho^2 \bm{\rho}_\mu \cdot \bm{\rho}^\mu, \\
  \mathcal{L}_\delta &=& \frac{1}{2} \partial_\mu \bm{\delta} \cdot \partial^\mu \bm{\delta} - \frac{1}{2} m_\delta^2 \bm{\delta}^2.
\end{eqnarray}
The symbols of $\kappa_3$, $\kappa_4$, and $\xi_0$ are the self-interaction coupling constants. The symbols of $m_\sigma$, $m_\omega$, $m_\rho$, and $m_\delta$ are the masses of the corresponding mesons. The field tensors of $\omega^{\mu \nu}$ and $\rho^{\mu \nu}$ are respectively for the $\omega$ and $\rho$ mesons, which can be defined as $\omega^{\mu \nu} = \partial^\mu \omega^\nu - \partial^\nu \omega^\mu$, and $\bm{\rho}^{\mu \nu} = \partial^\mu \bm{\rho}^\nu - \partial^\nu \bm{\rho}^\mu - g_\rho (\bm{\rho}^\mu \times \bm{\rho}^\nu)$. The sixth term of nonlinear crossing interaction Lagrangian for $\sigma$, $\omega$ and $\rho$ mesons is expressed by
\begin{eqnarray}
  \label{eq:eqxenont4}
  \mathcal{L}_{\sigma \omega \rho} &=&
  \frac{\eta_1}{2M} g_\sigma m_\omega^2 \sigma \omega_\mu \omega^\mu
  + \frac{\eta_2}{4M^2} g_\sigma^2 m_\omega^2 \sigma^2 \omega_\mu \omega^\mu 
  + \frac{\eta_\rho}{2M} g_\sigma m_\rho^2 \sigma \bm{\rho}_\mu \cdot \bm{\rho}^\mu \nonumber \\
  &+& \frac{\eta_{1\rho}}{4 M^2} g_\sigma^2 m_\rho^2 \sigma^2 \bm{\rho}_\mu \cdot \bm{\rho}^\mu 
  + \frac{\eta_{2 \rho}}{4 M^2} g_\omega^2 m_\rho^2 \omega_\mu \omega^\mu \bm{\rho}_\mu \cdot \bm{\rho}^\mu ,
\end{eqnarray}
where $\eta_1$, $\eta_2$, $\eta_\rho$, $\eta_{1\rho}$ and $\eta_{2 \rho}$ are the crossed interaction coupling constants. The values of all coupling constants in the effective Lagrangian of the E-RMF model in Eqs.~(\ref{eq:eqxenont2})--(\ref{eq:eqxenont4}) are tabulated in Table~\ref{tab:model1}.

%%%TABLE 1%%%%%%%%%%%%%%%
\begin{table}[t]
\caption{The complete parameters of G3(M)~\cite{Hutauruk:2021cgi} that was constrained by the PREX II data~\cite{PREX:2021umo}. The nucleon mass in free space $M$ is 939 MeV and coupling constants are dimensionless. The $k_3$ is in unit of fm$^{-1}$, and  $f_\omega / 4 =$ 0.220, $f_\rho / 4 =$ 1.239, and $\beta_\sigma =$ -0.0087. 
}
\label{tab:model1}
\addtolength{\tabcolsep}{0.00pt}
\begin{tabular}{ccccccccc} 
\hline \hline
 $m_s / M$ & $m_\omega /M$ & $m_\rho /M$ & $m_\delta /M$ & $\beta_\omega$ & $g_s /4\pi$ & $g_\omega / 4 \pi$ & $g_\rho / 4 \pi$ & $g_\delta / 4 \pi$ \\[0.2em] 
\hline
0.559 
& 0.832
   & 0.820 
 & 1.043 
    & 0.782 
 & 0.923
 & 0.872 
 & 0.160 
& 2.606  
\\ \hline \hline
\end{tabular}
\end{table} 

%%%TABLE 1
\begin{table}[t]
%\caption{The complete parameter set of G3(M) that determined by readjusting to the PREX-2 data~\cite{Adhikhari21}. The nucleon mass $M_N$ is 939 MeV and all coupling constants are dimensionless. The unit of $k_3$ is in fm$^{-1}$.
%}
%\label{tab:model1}
\addtolength{\tabcolsep}{0.00pt}
\begin{tabular}{ccccccccccc} 
\hline \hline
  $k_3$  & $k_4$ & $\beta_\omega$ & $\xi_0$ & $\eta_1$ & $\eta_2$ & $\eta_\rho$ & $\eta_{1 \rho}$ & $\eta_{2\rho}$ & $\alpha_1$ & $\alpha_2$ \\[0.2em] 
\hline
 1.694  
 &-0.484
 & 1.010 
 & 0.424
 & 0.114
 & 0.645 
 & 0.000 
 & 18.257 
 & 2.000 0
 & -1.468 
 & 0.220 
\\ \hline \hline
\end{tabular}
\end{table} 

Although, in this work, I explicitly do not include the electron and muon contributions in the calculation of the ADCS and AMFP, for completeness of the model Lagrangian, here I also present the Lagrangian density for the electron and muon. The expression of the Lagrangian density for the electron and muon can be given by
\begin{eqnarray}
  \mathcal{L}_l &=& \sum_{l =e,\mu} \bar{\psi}_l (i\gamma_\mu \partial^\mu - m_l) \psi_l,
\end{eqnarray}
with $m_l$ as the lepton mass in free space. It is with noting that the electron and muon masses do not change in nuclear medium. Using the given Lagrangian, the particle population or fractions of the NS standard matter with $n$, $p$, $e$, and $\mu$ can be calculated \textit{via} the $\beta$--equilibrium constraint that shows a relation of the chemical potential as
\begin{eqnarray}
  \mu_n - \mu_p &=& \mu_e, \hspace{1.0cm} \mu_e = \mu_\mu,
\end{eqnarray}
and the charge neutrality must be satisfied, which is given by
\begin{eqnarray}
  n_p &=& n_e + n_\mu,
\end{eqnarray}
where, in this case, the leptons are assumed as relativistic ideal Fermi gases. The chemical potential for protons and neutrons can be obtained through the equation $\mu_{p,n} = \frac{\partial \mathcal{E}}{\rho_{p,n}}$, and the $n_p$, $n_e$ and $n_\mu$ stand for the proton, electron, and muon densities, respectively. Because the lepton masses do not change with respect to the baryon number density, Then, the chemical potentials for lepton can be calculated by $\mu_{l=e,\mu} = \sqrt{k_{F_{l=e,\mu}}^2 + m_{l=e,\mu}^2}$. Note that the total baryon number density is defined by $n_B = n_p + n_n$ with $n_n$ as neutron number density. 

As explained earlier, the G3(M) parameter was well constrained by terrestrial experimental PREX II data. The difference between G3(M) and other parameter sets of the E-RMF model is that the G3(M) parameter has nonzero $\eta_1$, $\eta_2$, $\eta_\rho$, $\alpha_1$, $\alpha_2$, $f_\omega$, $f_\rho$, $\beta_\sigma$, and $\beta_\omega$ coupling constants, while other E-RMF parameter sets such the well-known TM1e and FSU Garnet parameter~\cite{Hutauruk:2021cgi} are set those coupling constants equal to zero. The complete G3(M) parameters are depicted in Table~\ref{tab:model1}. 

Instead of the difference in those coupling constants, the value of the $g_\delta$ coupling constant is also different. This coupling constant contributes to stiffening the EoS in the high density and the nuclear symmetry energy in the sub-saturation density. It is with noting that the nonlinear crossing coupling constants play a crucial role in obtaining a better EoS for PNM. This will affect the particle populations/fractions, ADCS, and AMFP. Note that the nonlinear crossing coupling constant $\eta_{2\rho}$ and the coupling constant $g_\rho$ in the G3(M) parameter are determined by fine-tuning $\eta_{2\rho}$ and $g_\rho$ parameters to reproduce the neutron skin thickness of ${}^{208}$Pb from the PREX II data, making a crucial different between the G3(M) parameter and other E-RMF parameter sets.

\begin{figure}[t]
\centering
\includegraphics[width=8.5cm]{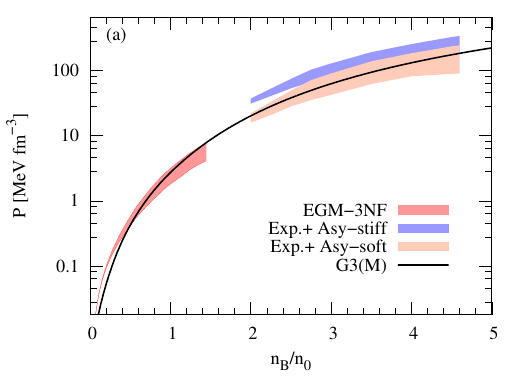}
\includegraphics[width=8.5cm]{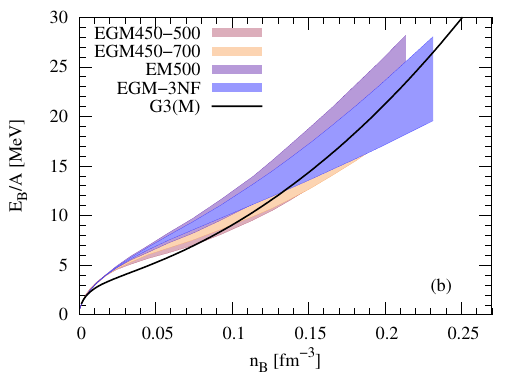}
\includegraphics[width=8.5cm]{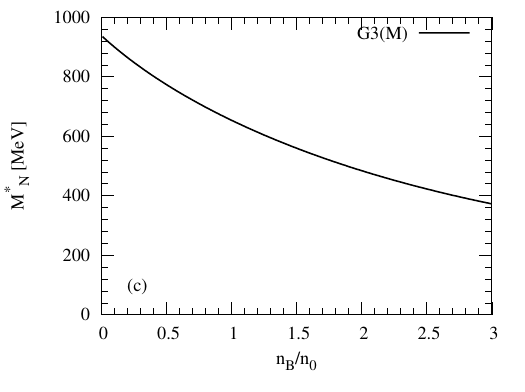}
\includegraphics[width=8.5cm]{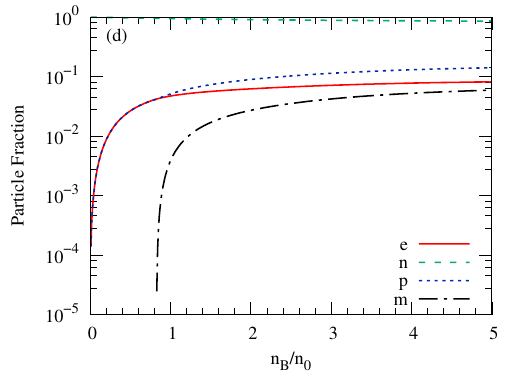}
\caption{(a) Pressure versus $n_B/n_0$ for G3(M) parameter, (b) Binding energy of a nucleon as a function of $n_B$, (c) Nucleon effective mass as a function of $n_B/n_0$, and (d) Particle fractions as a function of $n_B/n_0$..}
\label{fig1}
\end{figure}

Employing the effective Lagrangian in Eq.~(\ref{eq:eqxenont1}), the energy density $\mathcal{E}$ and pressure $P$ for NM can be evaluated using the energy-momentum tensor
 $ T_{\mu \nu} = \sum_k \partial_\nu \phi_k \frac{\partial \mathcal{L}}{\partial \left( \partial^\mu \phi_k\right)} - g_{\mu \nu} \mathcal{L}$
where $\phi_k$ are all corresponding fields in the effective Lagrangian of the E-RMF model in Eq.~(\ref{eq:eqxenont1}). Straightforwardly, the energy density will be obtained by calculating the zeroth component of the energy-momentum tensor $\langle T_{00} \rangle$ that gives $\mathcal{E} = \langle T_{00} \rangle$, and the pressure is obtained from the third component of the energy-momentum tensor $\langle T_{kk} \rangle$ that gives $ P = \sum_{k}\frac{1}{3} \langle T_{kk} \rangle $. The final expressions for the energy density and pressure expressions for PNM are respectively given by ~\cite{Furnstahl:1995zb,Furnstahl:1996wv}
\begin{eqnarray}
  \label{eq:edenPNM}
  \mathcal{E} &=& \sum_{i=n,p} \frac{2}{(2\pi)^3} \int_0^{k_F^i} d^3 k \sqrt{\bm{k}_i^2 + M_i^{* 2}} + n_B g_\omega \omega - \frac{1}{24} \xi_0 g_\omega^2 \omega^4 \nonumber \\
  &+& \frac{1}{2} m_\sigma^2 \sigma^2 \left( 1 + \frac{\kappa_3}{3M} g_\sigma \sigma + \frac{\kappa_4}{12 M^2} g_\sigma^2 \sigma^2 \right) \nonumber \\
  &-& \frac{1}{2} m_\omega^2 \omega^2 \left( 1 + \frac{\eta_1}{M} g_\sigma \sigma + \frac{\eta_2}{2M^2} g_\sigma^2 \sigma^2 \right) \nonumber \\
  &-& \frac{1}{2} m_\rho^2 \rho^2 \left( 1 + \frac{\eta_\rho}{M} g_\sigma \sigma + \frac{\eta_{1 \rho}}{2M^2} g_\sigma^2 \sigma^2 \right)- \frac{\eta_{2 \rho}}{2} g_\rho^2 g_\omega^2 \rho^2 \omega^2 \nonumber \\
  &+& \frac{1}{2} \rho_3 g_\rho \rho + \frac{1}{2} m_\delta^2 \delta^2,
\end{eqnarray}
and
\begin{eqnarray}
  \label{eq:pressPNM}
  P &=& \sum_{i =n,p} \frac{1}{3} \frac{2}{(2\pi)^3} \int_0^{k_F^i} d^3 k \frac{\bm{k}_i^2}{\sqrt{\bm{k}_i^2 + M_i^2}} + \frac{1}{24} \xi_0 g_\omega^2 \omega^4 \nonumber\\
  &-& \frac{1}{2} m_\sigma^2 \sigma^2 \left( 1 + \frac{\kappa_3}{3M} g_\sigma \sigma + \frac{\kappa_4}{12 M^2}g_\sigma^2 \sigma^2 \right) \nonumber \\
  &+& \frac{1}{2} m_\omega^2 \omega^2 \left( 1 + \frac{\eta_1}{M} g_\sigma \sigma + \frac{\eta_2}{2M^2} g_\sigma^2 \sigma^2 \right)\nonumber \\
  &+& \frac{1}{2} m_\rho^2 \rho^2 \left( 1 + \frac{\eta_\rho}{M} g_\sigma \sigma + \frac{\eta_{1 \rho}}{2M^2} g_\sigma^2 \sigma^2 \right) +  \frac{\eta_{2 \rho}}{2} g_\rho^2 g_\omega^2 \rho^2 \omega^2 \nonumber \\
  &-& \frac{1}{2} m_\delta^2 \delta^2, 
\end{eqnarray}
where the Fermi momentum $k_F$ can be defined from the baryon number density $n_B = \frac{k_F^3}{3\pi^2}$, and the isospin baryon density can be defined by $n_3 = n_p - n_n$. Numerical results of the E-RMF model with the G3(M) parameter for the pressure, the binding energy per nucleon, the effective nucleon mass, and the particle fractions are shown in Fig.~\ref{fig1}.

In Fig.~\ref{fig1}(a), I show the result for the pressure of the G3(M) parameter in comparison with the results of the chiral perturbation theory (ChPT)~\cite{Tews:2012fj} and heavy ion collision (HIC)~\cite{Danielewicz:2002pu}. Results for a nucleon's pressure and binding energy fit well with the other calculations~\cite{Tews:2012fj,Danielewicz:2002pu}. Results for the nucleon effective mass and the particle population fractions for G3(M) parameter are depicted in Figs.~\ref{fig1}(c) and (d), respectively.
%%%%

%-------------------------------------------------------------------
\subsubsection{KIDS-EDF model}
\label{sec:KIDSEDF}
%-------------------------------------------------------------------
In this section, I describe the KIDS-EDF model~\cite{Papakonstantinou:2016zpe,Hutauruk:2022bso}. The KIDS-EDF model has widely been applied for homogeneous NM~\cite{Papakonstantinou:2016zpe}, and finite nuclei. The EDF-KIDS model in the NM was transformed into the Skyrme functional form in finite nuclei, where the specific values of the effective masses could be emulated by adjusting the functional parameters. The energy of a nucleon for the KIDS-EDF model was described in the homogeneous NM as an expansion of the power of the Fermi momentum $k_F$ at zero temperature. The expression of the KIDS-EDF model energy per nucleon at zero temperature is explicitly given by ~\cite{Papakonstantinou:2016zpe,Hutauruk:2022bso}.
%%%%%%%%%%%%%%%%%%%%%
\begin{eqnarray}
\label{eqkids1}
\mathcal{E} (n_B,\delta) = \mathcal{T} (n_B, \delta) + \sum_{j=0}^{3} c_j (\delta) n_B^{[1+a_j]},
\end{eqnarray}
where $n_B = n_n + n_p$ is the baryon number density with $n_n$ and $n_p$ are the neutron and proton densities, respectively. The quantities of $a_j = j/3$, and $\delta = (n_n-n_p)/n_B$ is the isospin asymmetry parameter. The $\delta =0$ is for symmetric NM and $\delta =1$ is for pure NM. The kinetic energy of the first term in Eq.~(\ref{eqkids1}) can be defined by
\begin{eqnarray}
\mathcal{T} (n_B, \delta) = \frac{3}{5} (3\pi^2 n_B)^{\frac{2}{3}}\Bigg[ \frac{h}{4\pi M_p} \left( \frac{1-\delta}{2}\right)^{\frac{5}{3}}+ \frac{h}{4\pi M_n} \left( \frac{1+\delta}{2}\right)^{\frac{5}{3}}\Bigg],
\end{eqnarray}
where $M_n$ and $M_p$ stand for the neutron and proton masses in free space, respectively. It is worth noting that the variable of $c_j(\delta)$ in the potential term of Eq.~(\ref{eqkids1}) can be defined by $c_j (\delta) = \alpha_j + \delta^2 \beta_j$. The parameter of $\alpha_j$ can be fixed using the symmetry NM properties and $\beta_j$ can be fixed using the EoS of the asymmetric NM.

In determining the KIDS-EDF parameters based on the Skyrme force parameters, one is needed to match between them. Before explaining the matching parameter procedure, here, I shortly describe the conventional Skyrme forces interaction~\cite{Chabanat:1997un}
\begin{eqnarray}
  \label{eq2a}
  \mathcal{V}_{i,j} (\mathbf{k},\mathbf{k}') &=& t_0 (1 + x_0 P_\sigma ) \delta ( \mathbf{r}_i - \mathbf{r}_j) + \frac{1}{2} t_1 (1 + x_1 P_\sigma ) \left[ \delta (\mathbf{r}_i - \mathbf{r}_j)\mathbf{k}^2 + \mathbf{k}^{'2} \delta (\mathbf{r}_i - \mathbf{r}_j )\right] \nonumber \\
  &+& t_2 (1 + x_2 P_\sigma ) \mathbf{k}' \cdot \delta (\mathbf{r}_i - \mathbf{r}_j ) \mathbf{k} + \frac{1}{6} t_{3} (1 + x_3 P_\sigma) \rho^{\alpha} \delta (\mathbf{r}_i - \mathbf{r}_j) \nonumber \\
  &+& i W_0 \mathbf{k}' \times \delta (\mathbf{r}_i - \mathbf{r}_j)\mathbf{k} \cdot (\sigma_i - \sigma_j),
\end{eqnarray}
where the spin exchange operator can be defined by $P_\sigma = \left(1 + \sigma_1 \cdot \sigma_2 \right)/2$ with $\sigma$ are the Pauli spin matrices. The relative momenta operating in the initial and final states are given by
$\mathbf{k} = (\nabla_i - \nabla_j) /2i$ and $\mathbf{k}' = (\nabla_i^{'} - \nabla_j^{'}) /2i$ , respectively.
It is worth noting that the strength of the spin-orbit coupling $W_0$ is not considered in the EDF for homogeneous NM in the Skyrme force. The energy density for infinite NM in Eq.~(\ref{eqkids1}) can be decomposed in terms of the Skyrme parameters as
\begin{eqnarray}
  \label{eq2b}
  \mathcal{E} (n_B, \delta) &=& \mathcal{T} (n_B, \delta) + \frac{3}{8} t_0 n_B - \frac{1}{8} (2y_0 + t_0) n_B \delta^2 + \frac{1}{16} t_{3} n_B^{(\alpha +1)} \nonumber \\
  &-& \frac{1}{48} (2y_{3} + t_{3} ) n_B^{(\alpha +1)} \delta^2 + \frac{1}{16} (3t_1 + 5 t_2 + 4 y_2) \tau \nonumber \\
  &-& \frac{1}{16} \left[ (2 y_1 + t_1) - (2y_2 + t_2 )\right] \tau \delta^2,
\end{eqnarray}
where the definition of variable $y_i \equiv t_i x_i$ with $i=1,2,3$. It can be easily seen that matching Eqs.~(\ref{eqkids1}) and~(\ref{eq2b}) can be used to determine the relations between KIDS parameters $c_j(\delta)$ in terms of the Skyrme coefficients ($t_i,y_i)$ and one gives
\begin{eqnarray}
  \label{eq2c}
  c_0 (\delta) &=& \frac{3}{8} t_0 - \frac{1}{8} (2y_0 + t_0) \delta^2, \nonumber \\
  c_1 (\delta) &=& \frac{1}{16} t_{31} - \frac{1}{48} (2 y_{31} + t_{31} ) \delta^2, \nonumber \\
  c_2 (\delta) &=& \frac{1}{16} t_{32} - \frac{1}{48} (2 y_{32}+ t_{32}) \delta^2 \nonumber \\
&+& \frac{3}{5} \left( \frac{6\pi^2}{\nu} \right)^{\frac{2}{3}} \frac{1}{16} \left\{ (3t_1 + 5t_2 + 4y_2) -\left[(2y_1 + t_1) -(2y_2 + t_2)\right] \delta^2 \right\}, \nonumber \\
  c_3 (\delta) &=& \frac{1}{16} t_{33}- \frac{1}{48} (2y_{33}+ t_{33}) \delta^2,
\end{eqnarray}
where the variable $\alpha = 1/3$ is assigned to $t_{31}$, $y_{31}$, $\alpha = 2/3$ is assigned to $t_{32}$ and $y_{32}$, and $\alpha =1$ is assigned to $t_{33}$ and $y_{33}$. The spin and isospin degeneracy factor with $\nu=4$ is for symmetric NM and $\nu=2$ is the spin and isospin degeneracy for the asymmetric NM. It is worth noting that the KIDS-EDF model commonly has $13$ parameters in total to be determined in the Skyrme force. In the present work, I consider the various KIDS-EDF models: KIDS0, KIDSA, and SLy4 models.

Here, I describe specifically how to determine the parameters in the KIDS0 model. The parameters of $\alpha_0$, $\alpha_1$, and $\alpha_2$ have been determined to adjust to three symmetric NM data at saturation density $\rho_0=0.16$~fm$^{-3}$, binding energy $E_{\rm B} = 16$~MeV, and compressibility $K_0 = 240$~MeV. Four parameters of $\beta_i$'s are fitted to a PNM EoS. We then obtain the isoscalar effective mass $m^*_s \simeq 1.0 M$ and the isovector one $m^*_v \simeq 0.8 M$ as a result of the parameter fitting for the KIDS0 model. 

\begin{figure}[t]
\centering
\includegraphics[width=8.5cm]{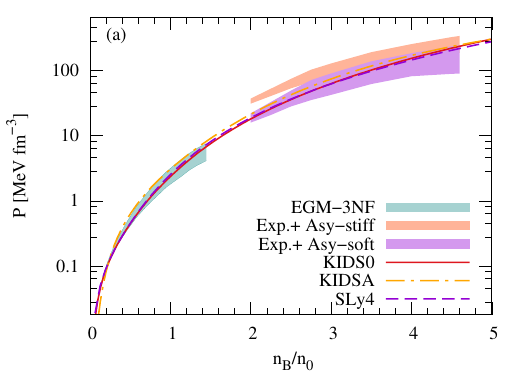}
\includegraphics[width=8.5cm]{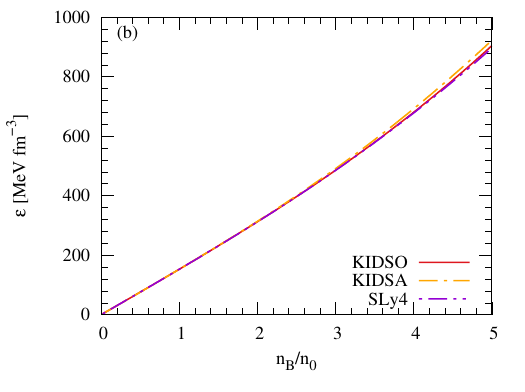}
\includegraphics[width=8.5cm]{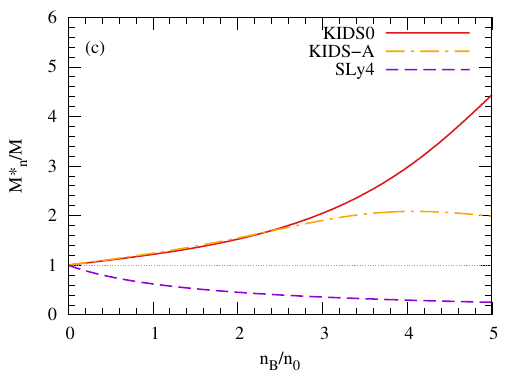}
\includegraphics[width=8.5cm]{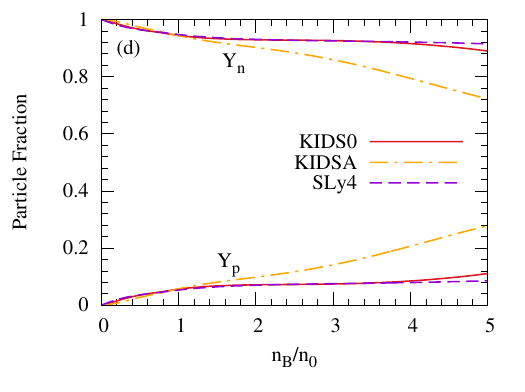}
\caption{(a) Pressure versus $n_B/n_0$ for the KIDS-EDF and SLy4 models, (b) Energy density versus $n_B/n_0$, (c) Neutron effective mass as a function of $n_B/n_0$, and (d) Particle fractions as a function of $n_B/n_0$.}
\label{fig2}
\end{figure}

The nuclear symmetry energy for the KIDSA model is determined to reproduce the recent constraint of the NS observation. The nuclear symmetry energy parameters are consistent with the NS data within an error bar uncertainty. Note that the values of the effective masses in the KIDSA model are obtained quite similar to the isoscalar and isovector effective masses for the KIDS0 model. The difference between the KIDS0 and KIDSA models is clearly shown in the nuclear symmetry energy. However, the KIDS0 and SLy4 models have similar results on the nuclear symmetry energy. Their difference is obviously indicated by the effective nucleon masses, as shown in Figs.~\ref{fig2}(c) and (d).

Based on the standard Skyrme force matching with the KIDS-EDF model parameters~\cite{Chabanat:1997un}, the effective nucleon masses in the asymmetric NM can be defined in terms of the Skyrme parameters. It can be given\cite{Papakonstantinou:2016zpe,Hutauruk:2022bso}
\begin{eqnarray}
M^*_i = M_i \left[ 1 + \frac{M_i}{8 \hbar^2} \rho \Theta_s - \frac{M_i}{8 \hbar^2} \tau^i_3 \left( 2 \Theta_v - \Theta_s \right) \rho \delta \right]^{-1},
\label{eq:efmn}
\end{eqnarray}
where parameter of $M_i$ is the proton and neutron masses in free space ($i=n,\, p$). The parameters of the Skyrme force can be respectively defined by $\Theta_s = 3t_1 + (5t_2+4y_2)$ and $\Theta_v=(2 t_1 +y_1) + (2 t_2 +y_2)$. The values of $t_1$, $t_2$, $y_1$, and $y_2$ are provided in Table~\ref{tab1}. The third component of the isospin of a nucleon with $\tau_3 = +1, \, -1$ are for the neutron and proton, respectively. Having the parameters of the Skyrme force of the nucleon isoscalar $\Theta_s$ and isovector mass $\Theta_v$, their ratios can be respectively given by
\begin{eqnarray}
  \label{eq3a1}
  \mu_s^* =m^*_s/M= \left( 1 + \frac{M}{8 \hbar^2} \rho \Theta_s \right)^{-1}, \,\,\,\,
  \mu_v^* =m^*_v/M = \left( 1 + \frac{M}{4 \hbar^2} \rho \Theta_v \right)^{-1}.
\end{eqnarray}
Through the equation above, it can be easily seen that the effective nucleon masses can be written in terms of the isoscalar and isovector masses \textit{via} Eqs.~(\ref{eq:efmn}) and~(\ref{eq3a1}). The values of the Skyrme force parameters, isoscalar and isovector effective masses, $K_0$, $J$, and $L$ for the KIDS0, and KIDSA models are summarized in Table~\ref{tab1}.
%%%TABLE 1 %%%%%%%
\begin{table}[t]
  \caption{A complete parameters of the Skyrme force for the KIDS0 and KIDSA models~\cite{Hutauruk:2022bii}. 
The units of $t_0, y_0$ are in MeV fm$^{3}$, the units of $t_{31}, y_{31}$ are in MeV fm$^{4}$, the units of $t_1$, $t_2$, $t_{32}$, $y_{32}$, $W_0$ are in MeV fm$^{5}$, and the unit of $y_{33}$ is in MeV fm$^{6}$. $K_0$, $J$, and $L$ are in units of MeV.}
  \label{tab1}
  \addtolength{\tabcolsep}{6.5pt}
  \begin{tabular}{ccccc} 
    \hline \hline
    Parameters & KIDS0 & KIDSA   \\[0.2em] 
    \hline
    $t_0$   & $-1772.044$ & $-1855.377$ \\
    $y_0$   & $-127.524$  & 2182.404  \\
    $t_1$   &  275.724  & 276.058  \\
    $y_1$   &  0.000    & 0.000   \\
    $t_2$   & $-161.507$  &$-167.415$  \\
    $y_2$   & 0.000       & 0.000   \\
    $t_{31}$ & 12216.730 & 14058.746   \\
    $y_{31}$ & $-11969.990$ & $-73482.960$   \\
    $t_{32}$ & 571.074 & $ -1022.193$   \\
    $y_{32}$ & 29485.421 & 122670.831  \\
    $t_{33}$ & 0.000 & 0.000  \\
    $y_{33}$ &$ -22955.280$ & $ -73105.329$ \\
    $W_0$   & 108.359 & 92.023 \\ \hline
    $\mu^*_s$ & 0.991 & 1.004  \\
    $\mu^*_v$ & 0.819 & 0.827 \\ 
    $K_0$ & 240 & 230  \\
    $J$ & 32.8 & 33 \\
    $L$ & 49.1 & 66  \\ 
    \hline \hline
  \end{tabular}
\end{table}
Figure~\ref{fig2} shows the numerical results for the pressure, energy density, effective neutron mass, and particle fractions for the KIDS0, KIDSA, and SLy4 models. A comparison between the pressure results for the KIDS0, KIDSA, and SLy4 models and ChPT and HIC predictions is given in Fig.~\ref{fig2}(a).

%-------------------------------------------------------------------
\section{Neutrino-NS matter constituents scattering}
\label{sec:scatt}
%-------------------------------------------------------------------
In this section, I present a generic formalism for the neutrino and antineutrino interaction with NS matter via the neutral current (NC) scattering with free space nucleon form factor at zero temperature. In the literature, there is a calculation of DCS of neutrino that was performed in the weak and electromagnetic interaction including the neutrino form factors~\cite{Hutauruk:2020mhl},  however, to avoid a complication of the calculation, in this work, I limit the current study only focusing on the standard weak interaction. The relevant interaction Lagrangian for the neutrino and antineutrino NC scattering in terms of the current-current interaction is written by~\cite{Hutauruk:2021cgi,Reddy:1997yr}
%%%%%%%%%%
\begin{eqnarray}
    \mathcal{L}_{\mathrm{INT}}^{NC} &=& \frac{G_F}{\sqrt{2}} \left[ \bar{\nu}_e \gamma^\mu \left( 1-\gamma_5 \right) \nu_e \right] \left[ \bar{\psi} \Gamma_\mu^{[n,p],NC} \psi \right]
\end{eqnarray}
where $G_F = 1.023/M \times 10^{-5}$ is the weak coupling constant and for standard nucleon vertex (including free space form factor of a nucleon) is defined $\Gamma_\mu^{[n,p],NC} = \gamma_\mu \left(C_V^{[n,p]} - C_A^{[n,p]} \gamma_5 \right)$. The values of $C_V$ and $C_A$ for neutrons and protons can be found in Refs.~\cite{Reddy:1997yr,Hutauruk:2022bii,Hutauruk:2018cgu,Hutauruk:2010tn,Sulaksono:2006eu,Sulaksono:2005wv,Kalempouw-Williams:2005zbp}. It is worth noting that $C_V \equiv F_{1W}(Q^2=0)$ and $C_A \equiv G_A$, where $F_{1W} (Q^2=0)$ and $G_A$ are respectively weak vector nucleon form factor and axial vector coupling constant. The values of $G_A$ and $F_{1,2}^W$ at $q^2 =0$ in free space are provided in Table.~\ref{tab2}. Note that, for antineutrinos, the sign of $G_A$ will be replaced by $-G_A$, and $C_V \equiv F_{1W}$ will be kept the same.
%%%%%%%%%%%%Table 2%%%%%%%%%%%%%%%%
\begin{table}
  \begin{center} 
  \caption{Free space nucleon axial and vector form factor values at $q^2=0$. In this work, we use $\sin^2\theta_w = 0.231$, $g_A=1.260$, $\kappa_p=1.793$ and $\kappa_n=-1.913$.}
\label{tab2}    
  \begin{tabular}{ccc}\hline
    Target & $G_A$ & $ C_V = F^{\rm W}_1$ \\ \hline
    $n$ &$-\frac{g_A}{2}$  & $-0.5$ \\
    $p$ & $\frac{g_A}{2}$ & $0.5 - 2 \sin^2 \theta_w$  \\
    %$e$ & $0.5 + 2 \sin^2 \theta_w$ & $\frac{1}{2}$ & 0 & 1 & 0 \\
    %$\mu$ & $-0.5 + 2 \sin^2 \theta_w$ & $-\frac{1}{2}$ & 0 & 1 & 0 \\
    \hline
  \end{tabular}
\end{center}
\end{table}
%%%%%%%%%%%%%%%%%%%%%%%%%%%%%%%%

%-------------------------------------------------------------------
\subsection{Neutral current interaction}
%-------------------------------------------------------------------
As mentioned above, in this work, we consider the NC weak interaction where the interaction between neutrino and NS matter constituents through the neutral $Z$ gauge bosons with the assumption the momentum transfer value is less than the masses of the weak gauge bosons. A generic expression for the double differential cross-section per volume for the neutrino and antineutrino scattering at zero temperature is given by~\cite{Hutauruk:2021cgi,Reddy:1997yr}
%%%%%%%%%%%%%
\begin{eqnarray}
\frac{1}{V} \frac{d^3 \sigma}{d^2 \Omega' dE_{\nu}'} = - \frac{G_F^2}{32\pi^2} \frac{E_{\nu}'}{E_{\nu}} \mathrm{Im} \left[L_{\mu \nu} \Pi^{\mu \nu} \right], 
\end{eqnarray}  
%%%%%%%%%%%%%
 where $E_\nu$ and $E'_\nu = E_\nu -q_0$ are the initial and final neutrino energies. The leptonic and hadronic tensors are respectively given by
\begin{eqnarray}
    L_{\mu \nu} = 8\left[ 2 k_\mu k_\nu + (k.q) g){\mu \nu} - (k_\mu q_\nu + q_\mu k_\nu) \mp i \epsilon_{\mu \nu \alpha \beta k^\alpha q^\beta} \right],
\end{eqnarray}
where the sign in the last term of the leptonic tensor is minus ($-$) for neutrinos and plus ($+$) for antineutrinos.
and
\begin{eqnarray}
    \Pi_{\mu \nu}^{[n,p]} (q^2) = -i \int \frac{d^4 p}{(2\pi)^4} \mathrm{Tr} \left[ G^{[n,p]} (p) \Gamma_\mu^{[n,p]} G^{[n,p]} (p+q) \Gamma_\nu^{[n,p]} \right], 
\end{eqnarray}
where the neutron and proton propagators in the nuclear medium can be defined by
\begin{eqnarray}
    G^{[n,p]} (p) = \Big[ \frac{p\!\!\!/^{*}+M^*}{p^{*2} - M^{*2} + i\epsilon} + i\pi \frac{p\!\!\!/^{*} + M^*}{E*} \delta \left( p_0^* -E^*\right) \Theta \left(P_F^{[n,p]} - |\mathbf{p}| \right)\Big].
\end{eqnarray}

After contracting the leptonic tensor and the polarization insertions (hadronic tensor) for the neutrons and protons, a final formula of neutrino and antineutrino DCS is given by~\cite{Hutauruk:2021cgi,Reddy:1997yr}
\begin{eqnarray}
   \frac{1}{V} \frac{d^3\sigma}{dE_\nu^{'} d^2 \Omega} = \frac{G_F^2}{4\pi^3} \frac{E_\nu^{'}}{E_\nu} q^2 \left[ A \mathcal{R}_1 + \mathcal{R}_2 + B \mathcal{R}_3 \right], 
\end{eqnarray}
where $A = \left[ 2E_\nu (E_\nu -q_0) + 0.5 q^2\right]/|\mathbf{q}|^2$, $B= 2E_\nu -q_0$, and $\mathcal{R}_1$, $\mathcal{R}_2$, and $\mathcal{R}_3$ are respectively given by
\begin{eqnarray}
    \mathcal{R}_1 &=& \left( C_V^2 + C_A^2 \right) \left[ \mathrm{Im} \Pi_L^{[n,p]} + \mathrm{Im} \Pi_T^{[n,p]} \right], \nonumber \\
    \mathcal{R}_2 &=& C_V^2 \Pi_T^{[n,p]} + C_A^2 \left[ \mathrm{Im} \Pi_T^{[n,p]} - \mathrm{Im} \Pi_A^{[n,p]} \right], \nonumber \\
    \mathcal{R}_3 &=& \pm 2 C_V C_A \mathrm{Im} \Pi_{VA}^{[n,p]},
\end{eqnarray}
where the plus ($+$) sign in $\mathcal{R}_3$ is for the neutrinos and the minus ($-$) sign in $\mathcal{R}_3$ is for the antineutrinos. In a nuclear medium, the polarization insertion can be decomposed into the polarization for the longitudinal, transversal, axial, and mixed vector-axial channels for the neutrons and protons respectively given by
\begin{eqnarray}
    \mathrm{Im} \Pi_L^{[n,p]} &=& \frac{q^2}{2\pi |\mathbf{q}|^3} \left[\frac{q^2}{4} \left( E_F-E* \right) + \frac{q_0}{2} \left( E_F^2 -E^{*2}\right) + \frac{1}{3} \left( E_F^3 - E^{*3}\right)\right],\\
    \mathrm{Im} \Pi_T^{[n,p]} &=& \frac{1}{4\pi |\mathbf{q}|} \Bigg[ \left( M^{*2} + \frac{q^4}{4|\mathbf{q}|^2} + \frac{q^2}{2} \right) \left( E_F -E^*\right) + \frac{q_0 q^2}{2|\mathbf{q}|^2} \left( E_F^2 -E^{*2}\right) \nonumber \\
    &+& \frac{q^2}{3|\mathbf{q}|^2} \left( E_F^3 - E^{*3} \right) \Bigg], \\
    \mathrm{Im} \Pi_A^{[n,p]} &=& \frac{i}{2\pi |\mathbf{q}|} M^{*2} \left( E_F - E^{*} \right),\\
    \mathrm{Im} \Pi_{VA}^{[n,p]} &=& \frac{iq^2}{8\pi |\mathbf{q}|^3} \left[ \left( E_F^2 - E^{*2} \right) + q_0 \left( E_F-E^*\right)\right].
\end{eqnarray}

The formula for the inverse neutrino and antineutrino MFP for the NC scattering at zero temperature is given by~\cite{Hutauruk:2021cgi,Reddy:1997yr}
\begin{eqnarray}
    \lambda^{-1} (E_\nu) = 2\pi \int_{q_0}^{(2E_\nu -q_0)} d|\mathbf{q}| \int_0^{2E_\nu} dq_0 \frac{|\mathbf{q}|}{E_\nu E_\nu'} \left[ \frac{1}{V} \frac{d^3 \sigma}{dE_\nu' d\Omega}\right]
\end{eqnarray}

%-------------------------------------------------------------------
\section{Numerical Results and Discussions}
\label{sec:result}
\begin{figure}[t]
\centering
\includegraphics[width=8.5cm]{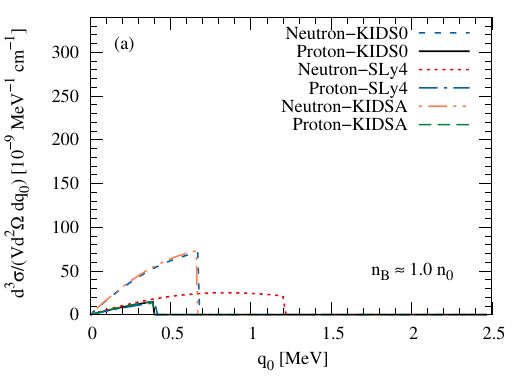}
\includegraphics[width=8.5cm]{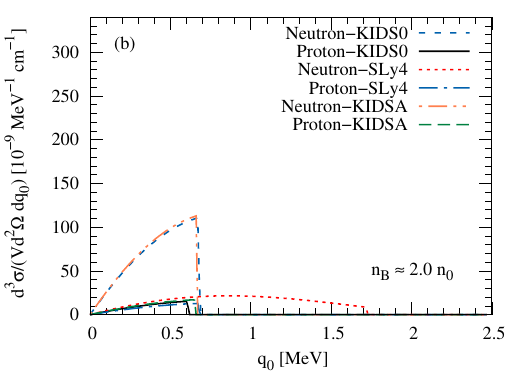}
\includegraphics[width=8.5cm]{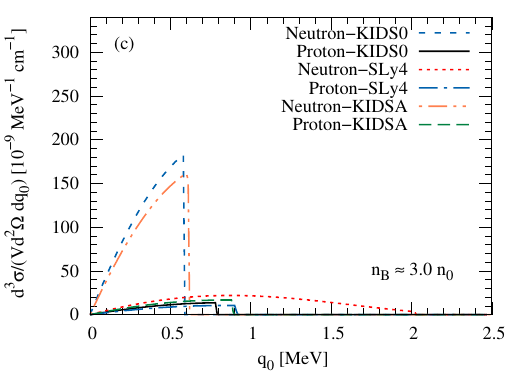}
\includegraphics[width=8.5cm]{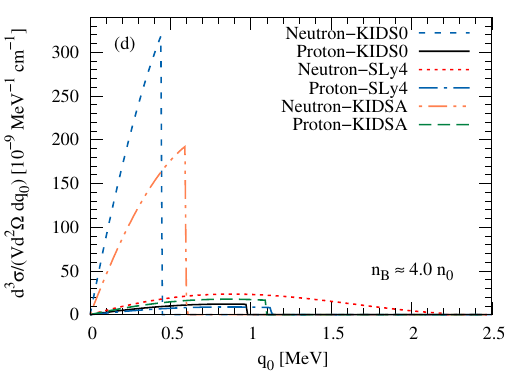}
\caption{Antineutrino differential cross-section for various KIDS-EDF: KIDSO, KIDSA, and SLy4 models that calculated at $E_\nu = $ 5 MeV and $|\textbf{q}|$ = 2.5 MeV with different densities (a) $n_B =$ 1.0 $n_0$, (b) $n_B =$ 2.0 $n_0$, (c) $n_B =$ 3.0 $n_0$, (d) $n_B =$ 4.0 $n_0$.}
\label{fig3}
\end{figure}

\begin{figure}[t]
\centering
\includegraphics[width=8.5cm]{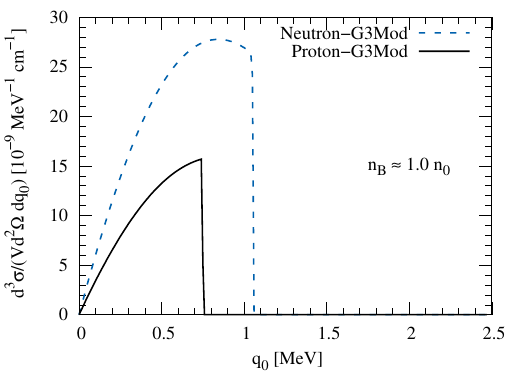}
\includegraphics[width=8.5cm]{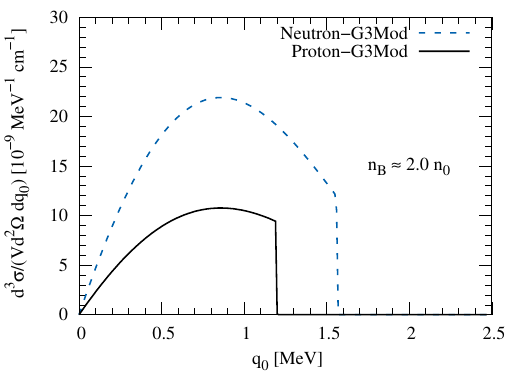}
\includegraphics[width=8.5cm]{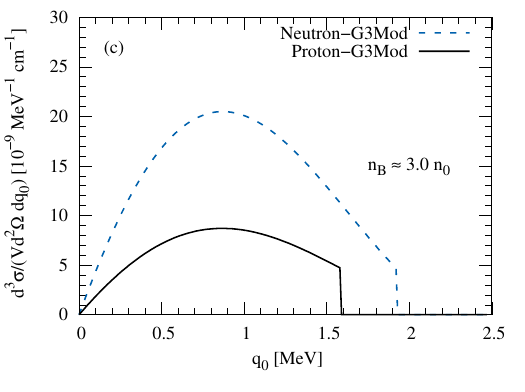}
\includegraphics[width=8.5cm]{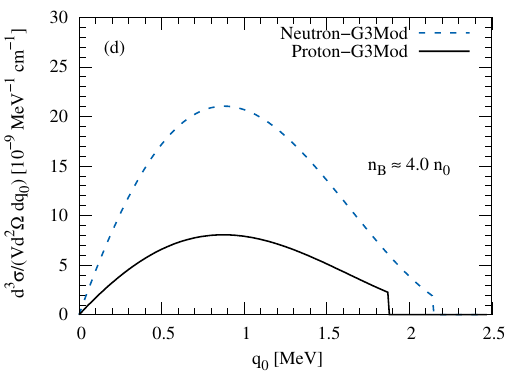}
\caption{Antineutrino differential cross-section for the E-RMF model with G3(M) parameter that calculated at $E_\nu = $ 5 MeV and $|\textbf{q}|$ = 2.5 MeV with different densities (a) $n_B =$ 1.0 $n_0$, (b) $n_B =$ 2.0 $n_0$, (c) $n_B =$ 3.0 $n_0$, (d) $n_B =$ 4.0 $n_0$.}
\label{fig4}
\end{figure}

\begin{figure}[t]
\centering
\includegraphics[width=12.5cm]{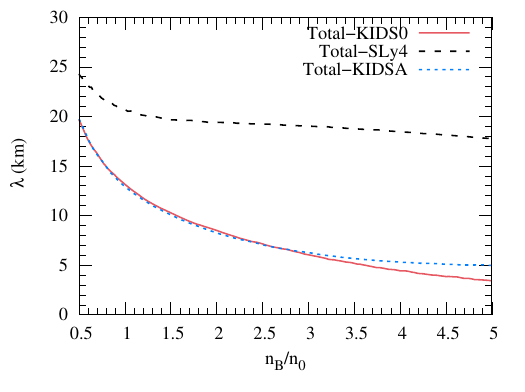}
\caption{Total AMFP for various KIDS-EDF: KIDS0, KIDSA and SLy4 models that calculated at $E_\nu = $ 5 MeV and $|\textbf{q}| =$ 2.5 MeV.}
\label{fig5}
\end{figure}

\begin{figure}[t]
\centering
\includegraphics[width=12.5cm]{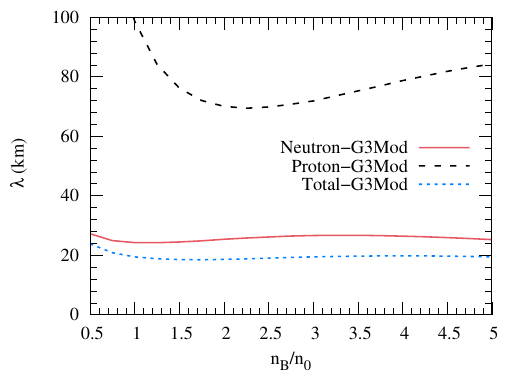}
\caption{Total and matter constituent contributions AMFP for the E-RMF model with G3(M) parameter that calculated at $E_\nu = $ 5 MeV and $|\textbf{q}|$ = 2.5 MeV.}
\label{fig6}
\end{figure}

%-------------------------------------------------------------------
Here, the numerical results for the antineutrino DCS and AMFP for the E-RMF, EDF-KIDS: KIDS0, KIDSA, and SLy4 models are presented in Figs.~\ref{fig3}-\ref{fig6}. Antineutrino DCS of neutrons and protons for the KIDS0, KIDSA, and SLy4 models at $n_b =$ 1.0 $n_0$ are shown in Fig.~\ref{fig2}(a). It shows the dominant contributions of the neutrons and protons scattering to the total DCS of antineutrino, which becomes a strong reason not to include the electron and muon scatterings in the present work, which gives a less contribution to the antineutrino DCS. One can conclude that the shape and magnitude of the antineutrino DCS strongly depends on the proton and neutron effective masses, as shown in Fig.~\ref{fig2}(c). Antineutrino DCS for neutron and proton of the KIDS0, KIDSA, and SLy4 models at different baryon densities $n_B =$ 2.0 $n_0$, $n_B=$ 3.0 $n_0$, and $n_B =$ 4.0 $n_0$ are respectively shown in Figs.~\ref{fig3}(a),(b), (c), and (d). They are calculated at the initial neutrino energy $E_\nu =$ 5 MeV and the three-component of momentum transfer $|\textbf{q}| =$ 2.5 MeV. Neutron contributions to the total antineutrino DCS for the KIDS0 and KIDS-A models are more pronounced at high baryon number density, as expected.

Similarly, the antineutrino DCSs of neutrons and protons for the E-RMF model with the G3(M) parameter are computed. The results of ADCS are shown in Fig.~\ref{fig4} with similar baryon number density, as depicted in Fig.~\ref{fig3}.
Figure~\ref{fig3}(a) shows that the antineutrino DCS of neutrons and protons for the E-RMF model with the G3(M) parameter at $n_B =$ 1.0 $n_0$. This result of the E-RMF model with the G3(M) parameter is almost similar in comparison to that for the SLy4 model. Again, it is obvious because of the effective masses of neutrons and protons, as their differences can be seen in Figs.~\ref{fig1}(c) and~\ref{fig2}(c). As the baryon number density increases, the range coverage of 
 $q_0$ becomes longer and the shape of the antineutrino DCS is rather changed as shown in Fig,~\ref{fig4}. However, the magnitude of the antineutrino DCS does not change significantly.

 Using the antineutrino DCS results for the E-RMF and EDF-KIDS models, the AMFP can be straightforwardly calculated for the corresponding models. The result of AMFP for the KIDS0, KIDSA, and SLy4 models is shown in Fig.~\ref{fig5}. It is clearly shown that the larger AMFP is found for the SLy4 model. For the KIDSO and KIDSA models, it shows that the AMFP for both models is the same up to around $n_B =$ 3.0 $n_0$ and it begins to diverge at $n_B >$ 3.5 $n_0$.

 AMFP results of neutrons and protons, as well as the total (neutrons + protons) for the E-RMF model with the G3(M) parameter, are depicted in Fig.~\ref{fig6}. Again, as antineutrino DCS for the SLy4 and E-RMF with the G3(M) parameter models, the antineutrino AMFP results for the SLy4 and E-RMF with the G3(M) parameter models are almost the same in magnitude, as expected. As conclusion is that the antineutrino DCS and AMFP strongly depend on the effective nucleon masses. The coverage range of $q_0$ is longer if the $q_0^{\rm{max}} > M_N^*$, which is related to the kinematic constraint of the nucleon.

%--------------------------------------------------
\section{Summary and Conclusion}
\label{sec:summary}
%--------------------------------------------------
In summary, we investigated the antineutrino DCS and MFP in the relativistic E-RMF model with the G3(M) parameter, inspired by an effective field theory (EFT), as well as in the nonrelativistic EDF-KIDS models: KIDS0, KIDSA, and the SLy4 model. To ensure the stability of neutron star matter, we also investigated the EoS for the G3(M), KIDS0, KIDSA, and SLy4 models to verify their reproduction of the binding energy per nucleon at saturation baryon number density.

From the present work, it was found that ADCS and AMFP strongly depend on the nucleon effective masses, and the range of q0 increases as the baryon number densityAstronomy 2024, 3 253 increases when the kinematic constraint $q_0^{\rm{max}} > M_N^*$ is fulfilled. This finding is similar to those found in the neutrino DCS and MFP.  

In qualitative comparison with the neutrino MFP, it was found that the antineutrino MFPis larger than the neutrino MFP, which is consistent with the result found in Ref. [5]. Even though the neutrino DCS and NMFP are not explicitly presented in this paper, the results for neutrino DCS and NMFP for the E-RMF and KIDS-EDF models are confirmed in Refs.~\cite{Hutauruk:2021cgi,Hutauruk:2022bii}.

This work also shows that the AMFP predictions of the E-RMF and KIDS-EDF models have rather different predictions on ADCS and AMFP, meaning the AMFP strongly depends on the nuclear models. This indicates the need for more data from terrestrial experiments and astrophysical observations to constrain the nuclear models better. Another approach would be for astrophysical experiments or observations (i.e., involving high-energy neutrino flux from the IceCube neutrino observatory) to directly measure the neutrino mean free path. This would be a great challenge and opportunity to compare empirical data with theory calculations directly.

%-------------------------------------------------
\section*{Acknowledgment}
%-------------------------------------------------
This work was partly supported by the National Research Foundation of Korea Grant Nos.~2018R1A5A1025563, 2022R1A2C1003964, 2022K2A9A1A0609176, and ~2023R1A2C1003177. The author thanks Profs. Seung-Il Nam (PKNU) and Chang-Ho Hyun (Daegu University) for the intriguing discussion and Prof. Anto Sulaksono (Indonesia University) for sharing the EoS code and the discussion.
%-------------------------------------------------

\appendix
\section{Abbreviations}
The abbreviations utilized in the paper are as follows:
\begin{table}[h]
  \addtolength{\tabcolsep}{6.5pt}
  \begin{tabular}{cc} 
    \hline 
    Abbreviations &  \\[0.2em] 
    \hline
    NC   & Neutral Current \\
    NS  & Neutron Star\\
    KIDS-EDF & Korea-IBS-Daegu-SKKU Energy Density Functional \\
    E-RMF   & Extended Relativistic Mean-Field   \\
    ADCS   &  Antineutrino Differential Cross-Section   \\
    AMFP  & Antineutrino Mean Free Path\\
    PNS &  ProtoNeutron Star \\
    NM & Nuclear Matter \\
    ChPT & Chiral Perturbation Theory  \\
    HIC &  Heavy Ion Collision \\
    NMFP &  Neutrino Mean Free Path\\
    \hline
  \end{tabular}
\end{table}

%-------------------------------------------------

%

\begin{thebibliography}{99}
\bibitem{LIGOScientific:2017vwq}
B.~P.~Abbott \textit{et al.} [LIGO Scientific and Virgo],
GW170817: Observation of Gravitational Waves from a Binary Neutron Star Inspiral,
Phys. Rev. Lett. \textbf{119}, no.16, 161101 (2017).

\bibitem{Hirata:1988ad}
K.~S.~Hirata, T.~Kajita, M.~Koshiba, M.~Nakahata, Y.~Oyama, N.~Sato, A.~Suzuki, M.~Takita, Y.~Totsuka and T.~Kifune, \textit{et al.}
Observation in the Kamiokande-II Detector of the Neutrino Burst from Supernova SN 1987a,
Phys. Rev. D \textbf{38}, 448-458 (1988).

\bibitem{Burrows:1987zz}
A.~Burrows and J.~M.~Lattimer,
Neutrinos from SN 1987A,
Astrophys. J. Lett. \textbf{318}, L63-L68 (1987).

\bibitem{IceCube:2016tpw}
M.~G.~Aartsen \textit{et al.} [IceCube],
All-sky Search for Time-integrated Neutrino Emission from Astrophysical Sources with 7 yr of IceCube Data,
Astrophys. J. \textbf{835}, no.2, 151 (2017).

\bibitem{Horowitz:2003yx}
C.~J.~Horowitz and M.~A.~Perez-Garcia,
Realistic neutrino opacities for supernova simulations with correlations and weak magnetism,
Phys. Rev. C \textbf{68}, 025803 (2003).

\bibitem{Hutauruk:2006re}
P.~T.~P.~Hutauruk, A.~Sulaksono and T.~Mart,
Effects of the neutrino electromagnetic form factors on the neutrino and antineutrino mean free paths in dense matter,
Nucl. Phys. A \textbf{782}, 400-405 (2007).

\bibitem{PREX:2021umo}
D.~Adhikari \textit{et al.} [PREX],
Accurate Determination of the Neutron Skin Thickness of $^{208}$Pb through Parity-Violation in Electron Scattering,
Phys. Rev. Lett. \textbf{126}, no.17, 172502 (2021).

\bibitem{Pattnaik:2021ido}
J.~A.~Pattnaik, R.~N.~Panda, M.~Bhuyan and S.~K.~Patra,
Constraining the relativistic mean-field models from PREX-2 data: effective forces revisited *,
Chin. Phys. C \textbf{46}, no.9, 094103 (2022).

\bibitem{Hutauruk:2021cgi}
P.~T.~P.~Hutauruk,
Implications of PREX-2 data on the electron-neutrino opacity in dense matter,
Phys. Rev. C \textbf{104}, no.6, 065802 (2021).

\bibitem{LIGOScientific:2018cki}
B.~P.~Abbott \textit{et al.} [LIGO Scientific and Virgo],
GW170817: Measurements of neutron star radii and equation of state,
Phys. Rev. Lett. \textbf{121}, no.16, 161101 (2018).

\bibitem{Miller:2019cac}
M.~C.~Miller, F.~K.~Lamb, A.~J.~Dittmann, S.~Bogdanov, Z.~Arzoumanian, K.~C.~Gendreau, S.~Guillot, A.~K.~Harding, W.~C.~G.~Ho and J.~M.~Lattimer, \textit{et al.}
PSR J0030+0451 Mass and Radius from $NICER$ Data and Implications for the Properties of Neutron Star Matter,
Astrophys. J. Lett. \textbf{887}, no.1, L24 (2019).

\bibitem{Riley:2019yda}
T.~E.~Riley, A.~L.~Watts, S.~Bogdanov, P.~S.~Ray, R.~M.~Ludlam, S.~Guillot, Z.~Arzoumanian, C.~L.~Baker, A.~V.~Bilous and D.~Chakrabarty, \textit{et al.}
A $NICER$ View of PSR J0030+0451: Millisecond Pulsar Parameter Estimation,
Astrophys. J. Lett. \textbf{887}, no.1, L21 (2019).

\bibitem{Papakonstantinou:2016zpe}
P.~Papakonstantinou, T.~S.~Park, Y.~Lim and C.~H.~Hyun,
Density dependence of the nuclear energy-density functional,
Phys. Rev. C \textbf{97}, no.1, 014312 (2018).

\bibitem{Mezzacappa:2000jb}
A.~Mezzacappa, M.~Liebendoerfer, O.~E.~B.~Messer, W.~R.~Hix, F.~K.~Thielemann and S.~W.~Bruenn,
The Simulation of a Spherically Symmetric Supernova of a 13 Solar Mass Star with Boltzmann Neutrino Transport, and its Implications for the Supernova Mechanism,
Phys. Rev. Lett. \textbf{86}, 1935-1938 (2001).

\bibitem{Rampp:2000ws}
M.~Rampp and H.~T.~Janka,
Spherically symmetric simulation with Boltzmann neutrino transport of core collapse and post-bounce evolution of a 15 solar mass star,
Astrophys. J. Lett. \textbf{539}, L33-L36 (2000).

\bibitem{Furnstahl:1996wv}
R.~J.~Furnstahl, B.~D.~Serot and H.~B.~Tang,
A Chiral effective Lagrangian for nuclei,
Nucl. Phys. A \textbf{615}, 441-482 (1997)
[erratum: Nucl. Phys. A \textbf{640}, 505-505 (1998)].

\bibitem{Furnstahl:1995zb}
R.~J.~Furnstahl, B.~D.~Serot and H.~B.~Tang,
Analysis of chiral mean field models for nuclei,
Nucl. Phys. A \textbf{598}, 539-582 (1996).

\bibitem{Hutauruk:2023mjj}
P.~T.~P.~Hutauruk, H.~Gil, S.~i.~Nam and C.~H.~Hyun,
Effects of Symmetry Energy on the Equation of State for Hybrid Neutron Stars,
[arXiv:2307.09038 [nucl-th]].

\bibitem{Hutauruk:2004uf}
P.~T.~P.~Hutauruk, C.~Kalempouw-Williams, A.~Sulaksono and T.~Mart,
Neutron fraction and neutrino mean free path predictions in relativistic mean field models,
Phys. Rev. C \textbf{70}, 068801 (2004).

\bibitem{Tews:2012fj}
I.~Tews, T.~Kr\"uger, K.~Hebeler and A.~Schwenk,
Neutron matter at next-to-next-to-next-to-leading order in chiral effective field theory,
Phys. Rev. Lett. \textbf{110}, no.3, 032504 (2013).

\bibitem{Danielewicz:2002pu}
P.~Danielewicz, R.~Lacey and W.~G.~Lynch,
Determination of the equation of state of dense matter,
Science \textbf{298}, 1592-1596 (2002).

\bibitem{Hutauruk:2022bso}
P.~T.~P.~Hutauruk, H.~Gil, S.~i.~Nam and C.~H.~Hyun,
Neutrino propagation in the neutron star with uncertainties from nuclear, hadron, and particle physics,
PTEP \textbf{2023}, no.6, 063D01 (2023).

\bibitem{Chabanat:1997un}
E.~Chabanat, P.~Bonche, P.~Haensel, J.~Meyer and R.~Schaeffer,
A Skyrme parametrization from subnuclear to neutron star densities. 2. Nuclei far from stabilities,
Nucl. Phys. A \textbf{635}, 231-256 (1998)
[erratum: Nucl. Phys. A \textbf{643}, 441-441 (1998)].

\bibitem{Hutauruk:2020mhl}
P.~T.~P.~Hutauruk, A.~Sulaksono and K.~Tsushima,
Effects of neutrino magnetic moment and charge radius constraints and medium modifications of the nucleon form factors on the neutrino mean free path in dense matter,
Nucl. Phys. A \textbf{1017}, 122356 (2022).

\bibitem{Reddy:1997yr}
S.~Reddy, M.~Prakash and J.~M.~Lattimer,
Neutrino interactions in hot and dense matter,
Phys. Rev. D \textbf{58}, 013009 (1998).

\bibitem{Hutauruk:2022bii}
P.~T.~P.~Hutauruk, H.~Gil, S.~i.~Nam and C.~H.~Hyun,
Effect of nucleon effective mass and symmetry energy on the neutrino mean free path in a neutron star,
Phys. Rev. C \textbf{106}, no.3, 035802 (2022).

\bibitem{Hutauruk:2019ptu}
P.~T.~P.~Hutauruk, Y.~Oh and K.~Tsushima,
Effects of medium modifications of nucleon form factors on neutrino scattering in dense matter,
JPS Conf. Proc. \textbf{26}, 024031 (2019).

\bibitem{Hutauruk:2018cgu}
P.~T.~P.~Hutauruk, Y.~Oh and K.~Tsushima,
The impact of medium modifications of the nucleon weak and electromagnetic form factors on the neutrino mean free path in dense matter,
Phys. Rev. D \textbf{98}, no.1, 013009 (2018).

\bibitem{Hutauruk:2010tn}
P.~T.~Hutauruk,
Neutrino Mean Free Path in Neutron Star,
[arXiv:1007.4007 [nucl-th]].

\bibitem{Sulaksono:2006eu}
A.~Sulaksono, C.~Kalempouw-Williams, P.~T.~P.~Hutauruk and T.~Mart,
Neutrino electromagnetic form factors effect on the neutrino cross section in dense matter,
Phys. Rev. C \textbf{73}, 025803 (2006).

\bibitem{Sulaksono:2005wv}
A.~Sulaksono, P.~T.~P.~Hutauruk and T.~Mart,
Isovector channel role of relativistic mean field models in the neutrino mean free path,
Phys. Rev. C \textbf{72}, 065801 (2005).

\bibitem{Kalempouw-Williams:2005zbp}
C.~Kalempouw-Williams, P.~T.~P.~Hutauruk, A.~Sulaksono and T.~Mart,
Neutrino electromagnetic form factor and oscillation effects on neutrino interaction with dense matter,
Phys. Rev. D \textbf{71}, 017303 (2005).


\end{thebibliography}
\end{document}